\documentclass[12pt]{article}
\usepackage{epsf} 
\usepackage{amsmath}
\usepackage{amssymb}

\usepackage{pstricks-add}

\begin{document}

\newcommand{\sheptitle}
{Q-branes}

\newcommand{\shepauthor}
{
Steven Abel$^{a }$\footnote[1]{s.a.abel@durham.ac.uk}, 
and Alex Kehagias$^{b }$\footnote[2]{kehagias@central.ntua.gr}}

\newcommand{\shepaddress}
{
$^a$ Institute for Particle Physics Phenomenology, University of Durham, DH1 3LE, UK\\
$^b$ Physics Dept., National Technical University, 157 80 Zografou, Athens, Greece}

\newcommand{\shepabstract}
{
Non-topological solitons (Q-balls) are discussed in some stringy settings.
Our main result is that the dielectric D-brane system of Myers admits non-abelian Q-ball solutions on their world-volume,
in which $N$ D$p$-branes relax to the standard dielectric form outside  the Q-ball, but 
assume a more diffuse configuration at its centre. We also consider 
how Q-balls behave in the bulk of extra-dimensional theories, or on wrapped branes.
We demonstrate that they carry Kaluza-Klein charge and possess a corresponding Kaluza-Klein tower of states 
just as normal particles, and we discuss surface energy effects by 
finding exact Q-ball solutions in models with a specific logarithmic potential. 
}

\begin{titlepage}
\begin{flushright}

\end{flushright}
\vspace{0.7in}
\begin{center}
{\LARGE{\bf \sheptitle}}
\bigskip \\ \vspace{0.3in} \shepauthor \\ \vspace{0.1in} \mbox{} \\ {\it \shepaddress} \\ 
\vspace{0.5in}
{\bf Abstract} \bigskip \end{center} \setcounter{page}{0}
\shepabstract
\vspace{0.5in}
\begin{flushleft}
\end{flushleft}
\end{titlepage}

\def\sspace{\baselineskip = .16in}
\def\dspace{\baselineskip = .30in}
\def\beq{\begin{equation}}
\def\eeq{\end{equation}}
\def\bea{\begin{eqnarray}}
\def\eea{\end{eqnarray}}
\def\bq{\begin{quote}}
\def\eq{\end{quote}}
\def\ra{\rightarrow}
\def\lra{\leftrightarrow}
\def\ups{\upsilon}
\def\bq{\begin{quote}}
\def\eq{\end{quote}}
\def\ra{\rightarrow}
\def\un{\underline}
\def\ov{\overline}
\def\ord{{\cal O}} 

\newcommand{\plb}[3]{{{\it Phys.~Lett.}~{\bf B#1} (#3) #2}}
\newcommand{\npb}[3]{{{\it Nucl.~Phys.}~{\bf B#1} (#3) #2}}
\newcommand{\prd}[3]{{{\it Phys.~Rev.}~{\bf D#1} (#3) #2}}
\newcommand{\pr}[3]{{{\it Physics~Reports}~{\bf #1} (#3) #2}}
\newcommand{\ptp}[3]{{{\it Prog.~Theor.~Phys.}~{\bf #1} (#3) #2}}
\newcommand{\ijmpa}[3]{{{\it Int.~J.~Mod.~Phys.}~{\bf A#1} (#3) #2}}
\newcommand{\prl}[3]{{{\it Phys.~Rev.~Lett.}~{\bf #1} (#3) #2}}
\newcommand{\hepph}[1]{{\tt hep-ph/#1}}
\newcommand{\hepth}[1]{{\tt hep-th/#1}}
\newcommand{\grqc}[1]{{\tt gr-qc/#1}} 
\newcommand{\leqsim}{\,\raisebox{-0.6ex}{$\buildrel < \over \sim$}\,}
\newcommand{\geqsim}{\,\raisebox{-0.6ex}{$\buildrel > \over \sim$}\,}
\newcommand{\be}{\begin{equation}}
\newcommand{\ee}{\end{equation}}
\newcommand{\ba}{\begin{eqnarray}}
\newcommand{\ea}{\end{eqnarray}}
\newcommand{\nn}{\nonumber}
\newcommand{\cf}{\mbox{{\em c.f.~}}}
\newcommand{\ie}{\mbox{{\em i.e.~}}}
\newcommand{\eg}{\mbox{{\em e.g.~}}}
\newcommand{\mpl}{\mbox{$M_{pl}$}}
\newcommand{\ol}[1]{\overline{#1}}
\def\bomega{{\boldsymbol \omega}}
\def\bPee{{\boldsymbol P}}
\def\gev{\,{\rm GeV }}
\def\tev{\,{\rm TeV }}
\def\dd{\mbox{d}}
\def\etal{\mbox{\it et al }}
\def\half{\frac{1}{2}}
\def\Tr{\mbox{Tr}}
\def\bra{\langle}
\def\ket{\rangle}
\def\lim{\mbox{{\bf L}} }
\def\nlim{\mbox{{\bf NL}} }
\def\sclim{\mbox{\tiny{\bf L}} }
\def\scnlim{\mbox{\tiny{\bf NL}} }
\def\nlimc{\mbox{{\bf NL$_{closed}$}} }
\def\nlimo{\mbox{{\bf NL$_{open}$}} }
\def\Vp{V_{\parallel}}
\def\Vt{V_{\perp}} 
\def\vp{V_{\parallel}}
\def\vt{V_{\perp}} 
\def\wp{W_{\parallel}}
\def\wt{W_{\perp}} 
\def\Rt{R_{\perp}}
\def\ep{\varepsilon} 
\def\hp{{\hat{\Phi}}} 
\newcommand{\smallfrac}[2]{\frac{\mbox{\small #1}}{\mbox{\small #2}}}

\def\CAG{{\cal A/\cal G}}           \def\CO{{\cal O}} \def\CZ{{\cal Z}}
\def\CA{{\cal A}} \def\CC{{\cal C}} \def\CF{{\cal F}} \def\CG{{\cal G}}
\def\CL{{\cal L}} \def\CH{{\cal H}} \def\CI{{\cal I}} \def\CU{{\cal U}}
\def\CB{{\cal B}} \def\CR{{\cal R}} \def\CD{{\cal D}} \def\CT{{\cal T}}
\def\CM{{\cal M}} \def\CP{{\cal P}}
\def\CN{{\cal N}} \def\CS{{\cal S}}

\newpage 

\section{Introduction and Background}

One of the most interesting features of field theories with 
conserved charges is the possibility of non-topological solitons, 
in particular Q-balls~\cite{leepang,coleman,Lee:1988ag,SCA,kusenko,AFK,Arai:2008kv}. 
Q-balls are localized field configurations that are stable simply because they are a more energy 
efficient way of holding charge than a collection of 
asymptotically free quanta. Such objects have been shown to occur very generally in field theory. Indeed the charge Q can be global (the simplest case) or local~\cite{Lee:1988ag,SCA} and the corresponding symmetry (which is spontaneously broken in their 
interior) can be either abelian or non-abelian.

For certain types of potential, 
the energy deficit or binding energy grows with charge, so that 
Q-balls can in principle be large macroscopic objects, and 
naturally there has been much interest in their cosmological implications and their impact on 
scenarios beyond the Standard Model~\cite{kusenko2,star,star2}. For example,   Q balls have been proposed as dark matter candidates \cite{kusenko-dm,Graham} in particular in gauge
mediated SUSY breaking models \cite{Kusenko:1997vp,Gia,Yanagida,japan2}. Experimental searches for Q-balls have also been proposed
and carried out. For these, 
the possible electric charge a Q-ball may carry obviously plays a central role in determining their experimental signature.  For instance,  neutral
Q-balls can be detected by Super-Kamiokande 
\cite{Kusenko:1997vp,arafuta,Takenaga} by probing proton absorption. Conversely charged Q-balls could be seen directly in detectors such as 
MACRO \cite{Kusenko:1997vp,Ambrosio}. 

Given this interest, it is important to determine the ubiquity of Q-balls in scenarios of physics at the most fundamental scales. 
In this paper we study Q-lumps in various stringy settings, including configurations with
 extra dimensions, namely charged bulks, and wrapped branes. 
Our main result, in section 5, is the explicit construction 
of stable Q-ball solutions on systems of D$p$-branes, in which  
the scalar fields in the solutions describe their displacements. It is 
well known that the D$p$-branes can be  
spread over a 2-sphere by turning on a background field, forming so-called 
dielectric branes~\cite{myers}.
The global minimum of a dielectric brane has a non-commutative 
form, with the vacuum falling into an $N\times N$
irreducible representation of $SU(2)$. However, in this minimum there are 
additional non-abelian symmetries that can be broken by reducible representations of 
$SU(2)$. We show that the resulting charges support Q-balls, with
the $N$ D$p$-branes relaxing to the standard dielectric form outside  the Q-ball, but 
assuming a more complicated dielectric configuration at its centre, in which the 2-sphere 
itself is diffuse.
Remarkably, even in the simplest case 
the dielectric brane potential has the correct 
coefficients for the Q-ball configuration to be energetically stable.

As well as presenting this construction we will, as a warm-up exercise,
look at a number of additional issues that make Q-balls in 
extra dimensional setups a somewhat more complex
problem than in 4 dimensional field theory.
The first is that generally they will 
be wrapped on compact dimensions of various size. 
The extent of the Q-ball can therefore be limited, 
forcing the configurations to be anisotropic. 
The second issue, is that the Q-balls carry a Kaluza-Klein 
momentum in the extra dimensions which is quantized. Thus 
one expects to find a tower of Q-balls, corresponding to Kaluza-Klein 
excitations. In the limit of large compactification, 
one naturally expects the momentum to become continuous
corresponding to the Q-balls moving freely in the extra-dimensions. 

We shall look at these issues by way of introduction to Q-balls 
in the following two sections, using a $U(1)$ model in 
5 space time dimensions with one compact space dimension
(\ie corresponding to a Q-ball on a wrapped 4-brane). 
We first discuss, in section 2, the large volume limit of Q-balls 
for complex fields carrying both global charge, $Q$, and Kaluza-Klein 
momentum of a single compact extra dimension, $P_5$. 
The solutions are found to have rather natural $O(3)$ and $O(4)$ symmetric limits 
depending on the size of the extra dimension. 
We find that the momentum modes correspond to 
an infinite set of Kaluza-Klein excitations
of the lowest lying Q-ball; 
the spectrum has a tower of $P_5$ momenta,
\be 
P_5= Q \left (p+\frac{n}{R} \right) \mbox{\hspace{1cm}$ n \in {\mathbb Z}$} \, ,
\ee
where $Q$ is the global charge, $p$ is the lowest mode and $R$ is the 
compactification radius. The states with non-zero $n$ can be thought of 
as Kaluza-Klein excitation of the lowest mode.
If $p=0$ the lowest mode is precisely the usual $D=4$ Q-ball, albeit possibly constrained 
by compact extra dimensions, while $p\neq 0$
corresponds to giving this state additional mass by the Scherk-Schwarz 
mechanism~\cite{ss}. 

In section 3 we discuss 
Q-balls in a special logarithmic potential that allows us to reduce the task of finding a Q-ball solution to the
canonical one-dimensional problem, showing in detail 
a Q-ball going from $O(4)$ to $O(3)$ symmetric configurations, and demonstrating the energetic preference of the surface tension 
term for large radii and more symmetric configurations. Finally in section 4 we discuss the non-abelian Q-ball solutions 
that are shown generally to exist on dielectric branes. 

\section{Warm up; large Q-balls in small boxes}

In order to see how Q-balls behave with finite dimensions, 
consider the large charge limit of a Q-ball in 5 space-time dimensions.
In this limit one neglects the surface effects. (In the
following section we discuss these  
using a particular logarithmic potential.)

The specific set-up is as follows. 
We shall take a single scalar field in 
$M_4 \times S^1 $. The Minkowski dimensions we call $x$,
and the dimension that is compactified on $S^1$ we call $y$, with 
$y$ and $y + 2\pi R$ identified. Almost certainly the discussion will hold also for
the untwisted sector of orbifolded extra dimensions, and as will become clear the 
qualitative behaviour would most likely be the same in non-flat compactifications. 

The action can be written as
\be 
S= \int \dd^3 x \dd y \dd t ~ \bigg(
\partial_\alpha \phi^* \partial^\alpha \phi - U_5 
(\phi \phi^*)\bigg).
\ee
Reparameterization invariance leads to the following 
conserved charges;
\ba
E &=& \int \dd ^3 x \dd y ~
\bigg(\partial_t \phi^* \partial_t \phi +
\partial_i \phi^* \partial_i \phi +
\partial_y \phi^* \partial_y \phi + 
U_5(\phi \phi^* ) \bigg)\nonumber \\
P_i &=& \int \dd ^3 x \dd y ~\bigg(\partial_t \phi^* \partial_i \phi +
\partial_i \phi^* \partial_t \phi \bigg)\nonumber \\
P_5 &=& \int\dd ^3 x \dd y ~ \bigg(\partial_t \phi^* \partial_y \phi +
\partial_y \phi^* \partial_t \phi \bigg)\, ,
\ea
where $i=1,2,3$. In addition, 
assume invariance under a global 
$U(1)$ transformation, $\phi \rightarrow 
e^{i \alpha } \phi $, so that there is a conserved charge,
\be 
Q=\frac{1}{i} \int \dd^3 x \dd y ~ 
\phi^* \stackrel{\leftrightarrow}{\partial}_t\phi .
\ee
By assumption the origin is a global minimum of $U_5$
and the global $U(1)$ symmetry is unbroken there.
As mentioned in the Introduction, in more general cases the transformation 
could be that of any compact group, and the Q-ball could be constructed from a 
local as well as a global symmetry. These extensions will be discussed in more detail 
later when we come to consider dielectric brane configurations.

Since we seek a solution that is localized in the $x$ coordinates
the global minimum in energy must have $U(1)$ symmetry restored 
at large radius for any $y$. Hence it is convenient to 
separate out the time dependent $U(1)$ phase;
\be 
\phi(x,y,t) = \varphi(x,y,t) e^{i \theta (y,t)},
\ee
where $\varphi$ and $\theta$ are real. The equations of 
motion now give us two relations, 
\ba 
\varphi \partial ^2 \theta + 2 \partial \theta \partial \varphi & = & 0 \nn\\
\partial ^2 \varphi - (\partial \theta)^2  \varphi & = & -\frac{1}{2} 
\frac{\partial U_5 }{\partial \varphi} \, .
\ea
By analogy with standard Q-balls we now choose $\theta $ to be linear,
parameterizing it with $\alpha $ and $\omega $;
\be 
\theta = \alpha (t+\omega y ) .
\ee
The equations of motion then require 
\be 
\varphi = \varphi ( y+\omega t ) 
\ee
and 
\be 
\label{EoM}
(1-\omega ^2 ) \partial _5^2 \varphi + \partial _i ^2 \varphi = \frac{1}{2} 
\frac{\partial \hat{U} }{\partial \varphi} 
\ee
where 
\be 
\hat{U} = U_5 - (1-\omega^2 ) \alpha^2 \varphi^2 .
\ee
The Q-ball solution then corresponds to the usual problem of 
a real field rolling in the inverted potential $-\hat{U}$ 
where $y$ and $x_i$ replace time. 

For completeness let us find the same result using the, 
perhaps more familiar, method which deduces the 
solution by minimising the energy of a generic field 
configuration whilst fixing the charges using 
Lagrange multipliers. That is we minimise the
expression~\cite{leepang}  
\ba
\label{eww}
\varepsilon _{\omega,\omega'} &=& E + \omega \left\{ 
P_5 - \int \dd ^3 x \dd y ~ \Big{(}\partial_t \phi^* \partial_y \phi +
\partial_y \phi^* \partial_t \phi \Big{)}\right\}
\nonumber \\
&& + \omega' \left(Q- \frac{1}{i} \int \dd^3  x \dd y ~
\phi^* \stackrel{\leftrightarrow}{\partial}_t\phi \right) 
\ea
for a given $\omega $ and $\omega'$, and then minimise in 
$\omega ,\omega'$.
First completing the square in the kinetic terms gives 
\ba
\label{first}
\varepsilon _{\omega,\omega'}
&=& \int \dd ^3 x \dd y ~\left(
\bigg|\omega \partial_y \phi - \partial_t \phi + 
i \omega' \phi \bigg|^2
+(1-\omega^2 )
\left|\partial_y \phi - \frac{i \omega \omega' }
{1-\omega^2}\phi \right|^2 \right)
\nonumber\\
&&
\hspace{0.75cm}+\int \dd ^3 x \dd y ~ 
\bigg(\partial_i \phi^* \partial_i \phi +
\hat{U}_{\omega\omega'}(\phi\phi^*) \bigg)\hspace{0.75cm}+\omega P_5 + \omega' Q\, ,
\ea
where
\be 
\hat{U}_{\omega\omega'}(\phi\phi^*)
=U_5(\phi\phi^* ) - \frac{{\omega'}^2 }{(1-\omega^2)}\phi\phi^*.
\ee
Now $\theta$ only appears in the first integral as
\ba
&&\int \dd ^3 x \dd y \,
\bigg(\bigg| i \varphi \left(\omega \partial_y\theta-\partial_t\theta+
\omega' \right))+ \omega\partial_y\varphi - \partial_t \varphi \bigg|^2
\nonumber \\
&&\hspace{2.0cm} +(1-\omega^2 )
\bigg||\partial_y \varphi - 
i\varphi \left( \partial_y \theta - \frac{ \omega \omega' }
{1-\omega^2} \right) \bigg|^2\bigg) \, .
\ea
The energy is minimised where the imaginary contributions vanish, 
which independently determines $\theta $,
\be 
\theta(y,t)= \frac{\omega'}{1-\omega^2} (\omega y + t ) ,
\ee
and in addition,
\be 
\varphi=\varphi (y+\omega t).
\ee
As one might have expected, the solutions with non-zero $P_5$ are going to be ``lumps'' travelling in the 
$y$ direction with speed $\omega$. The energy is now
\ba
\label{second}
\varepsilon _{\omega,\omega'}
&=& \int \dd ^3 x \dd y ~ 
\bigg((\partial_y \varphi)^2 +
(\partial_i \varphi)^2 +
\hat{U}_{\omega\omega'}(\varphi^2) \bigg)
\nonumber\\
&&\hspace{6.5cm}+\omega P_5 + \omega' Q ,
\ea
and clearly extremizing this gives the same equation of motion as before
if we identify $\alpha = \omega '/(1-\omega^2)$. The physical interpretation 
is that the boost factor (squared) $1/(1-\omega^2)$ is a result of both Lorentz contraction and time dilation in the 
phase factor, that will ultimately feed into the charge $Q$.  

We can now proceed to the large and small (in a sense to be defined shortly)
limits of compactification radius, $R$. In the large $R$ limit the 
$\varphi $ configuration that minimises $\varepsilon $ is 
approximately the same as that in the decompactified space.
The variation of $\varphi $ proceeds as for tunneling
in $d=4$ Euclidean space dimensions in the potential 
$\hat{U}_{\omega\omega'}$.
In this limit the symmetry of the problem dictates that, for the 
stationary ($\omega=0$) Q-ball, we
have a fully $O(4)$ symmetric solution and so 
the minimum is the action $S_4 [\varphi ]$
of the bounce solution. Since $\varphi$ is a function of $y+ \omega t $, 
the factor ($1-\omega^2 $) includes a Lorentz contraction 
which squashes the solution in the $y$ direction.
Clearly for the $O(4)$ limit to apply, $R$ should be 
much greater than the radius of this solution (which we shall call $r_4$).

\begin{figure}[t]
\begin{pspicture}[](0,0)(10,5)
\vspace{0.3in}
\hspace{0.3cm}
\epsfysize=2in
\epsffile{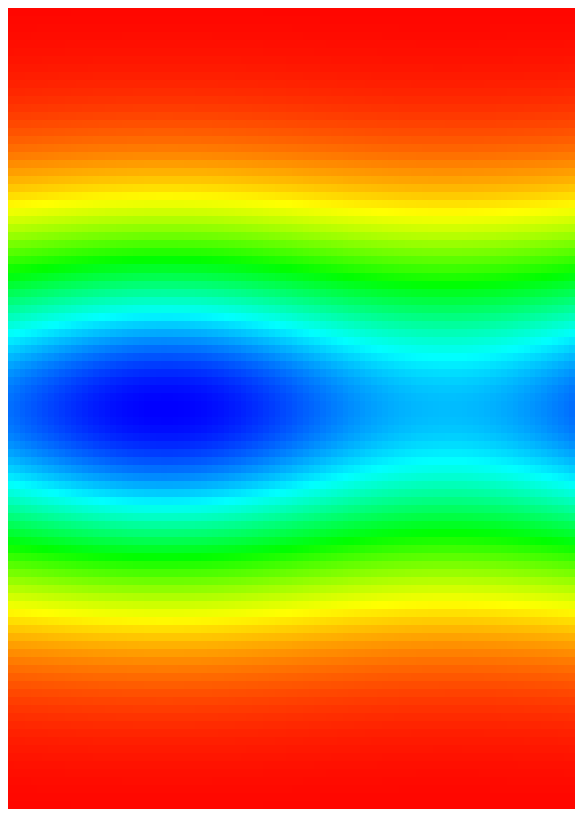}
\hspace{1.4cm}
\epsfysize=2in
\epsffile{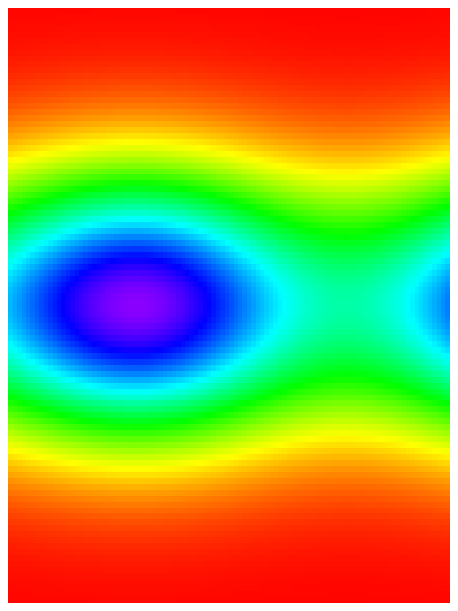}
\hspace{1.4cm}
\epsfysize=2in
\epsffile{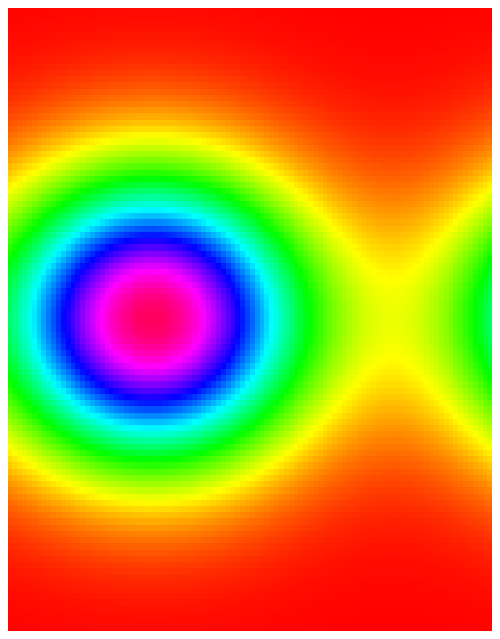}
\psline(0.01,0)(0.01,5.06)
\psline(-3.94,0)(-3.94,5.06)
\psline(-5.63,0)(-5.63,5.06)
\psline(-9.37,0)(-9.37,5.06)
\psline(-11.1,0)(-11.1,5.06)
\psline(-14.67,0)(-14.67,5.06)
\end{pspicture}
\\ \vspace{0.3in} \\
\begin{pspicture}[](0,0)(10,5)
\vspace{-1.3in}
\hspace{0.7cm}
\epsfysize=2in
\epsffile{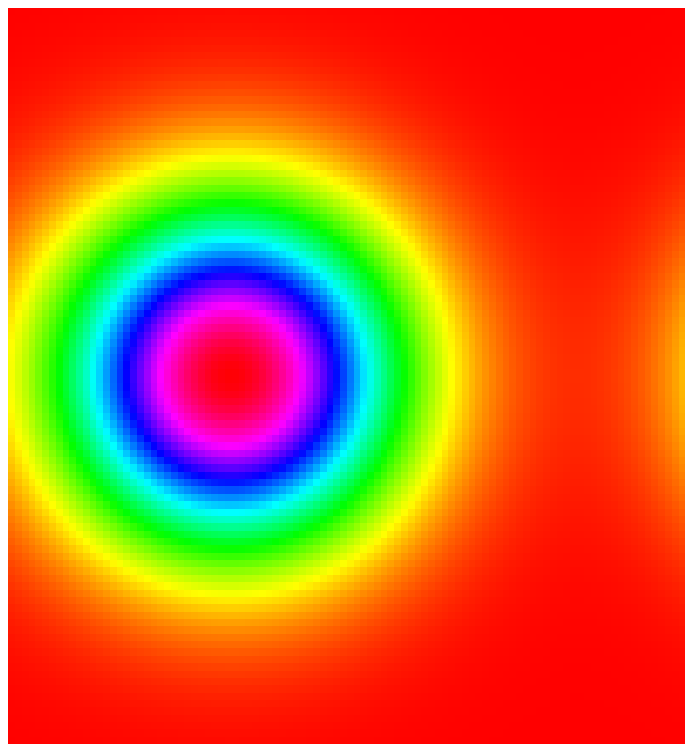}
\hspace{1.4cm}
\epsfysize=2in
\epsffile{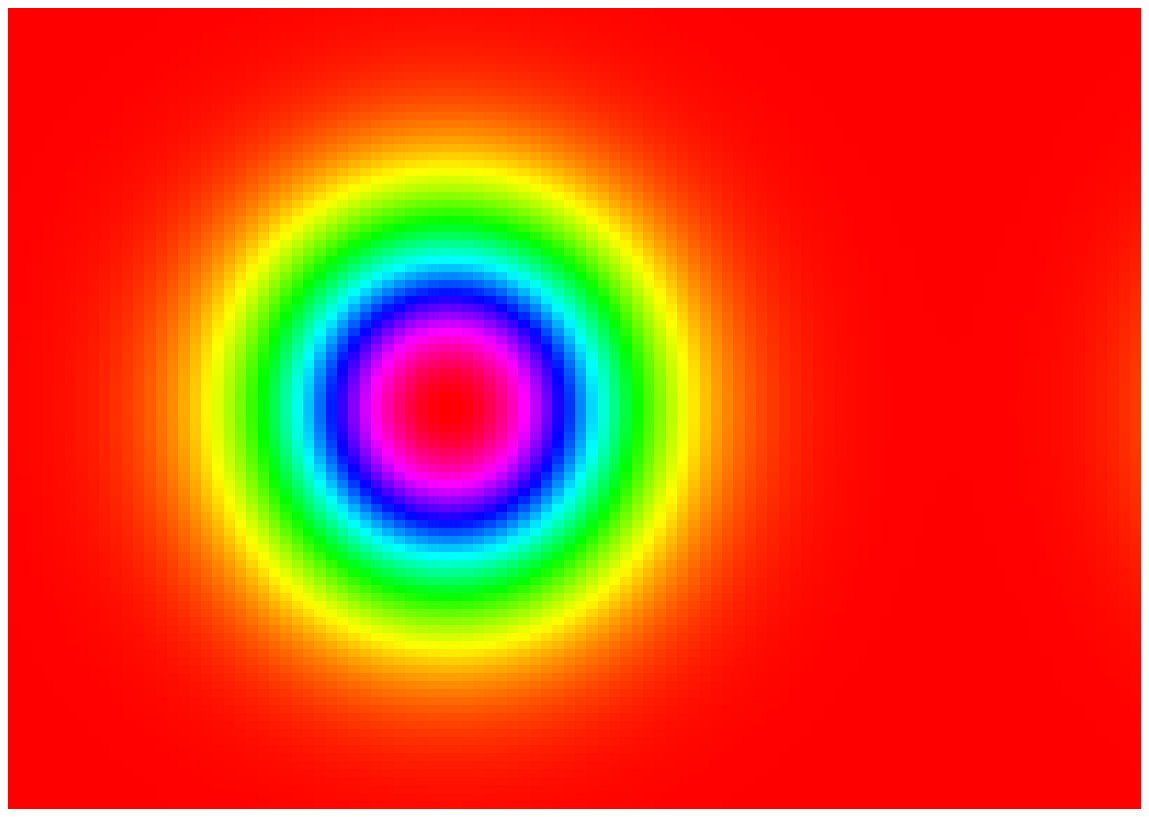}
\psline(0.01,0)(0.01,5.06)
\psline(-7.17,0)(-7.17,5.06)
\psline(-8.88,0)(-8.88,5.06)
\psline(-13.5,0)(-13.5,5.06)
\vspace{-3cm}
\rput(-3.05,5.8){c)}
\rput(-8.5,5.8){b)}
\rput(-13.75,5.8){a)}
\rput(-7.5,0.2){e)}
\rput(-13.85,0.2){d)}
\end{pspicture}
\caption{\it A Q-ball condensing as the bulk radius
increases. Here 
the centre of the Q-ball is offset by 1/2 a bulk radius. 
Figure 1e is close to the solution for an infinite radius. Figure 1a is slightly 
larger than the radius corresponding to the ``natural'' frequency of oscillation in the 
upturned potential. For a transverse radius smaller than a certain critical value, the only solution is trivial in the compact direction (i.e. constant in $y$). The slight 
squashing in the compactified direction is Lorentz contraction due to the non-zero $P_5$.} 
\end{figure}

In the opposite limit, where $R<r_4$, the $(\partial_y \varphi)^2 $ 
term makes any significant variation of $\varphi $
in the interval $y \in [0, 2 \pi R]$ very costly in energy. 
In this limit we can therefore take $\varphi (x,y) = \varphi(x)$
and write 
\ba
\label{third}
\varepsilon _{\omega,\omega'}
&=& 2 \pi R \int \dd ^3 x ~ 
\bigg((\partial_i \varphi)^2 +
\hat{U}_{\omega\omega'}(\varphi^2)\bigg)
\nonumber\\
&&\hspace{5cm}+\omega P_5 + \omega' Q .
\ea
Now the variation of $\varphi $ proceeds as for tunneling
in $d=3$ Euclidean dimensions in the potential 
$\hat{U}_{\omega\omega'}$, and the minimum energy is the action 
$S_3 [\varphi ]$ of the relevant bounce solution.
An example of the two limits, which we will be 
discussing in detail in the following sections, 
is shown in Figure 1.

Consider now the large volume solution, where the 
field is approximately constant, $\varphi_0$, 
inside a 4-volume $V_4$
(whose form depends on whether we are considering the 
large or small $R$ limit).
In this case we find 
\ba
Q &=& \frac{2\omega'}{1-\omega^2 } V_4 \varphi_0^2 \nonumber \\
P_5 &=& \frac{2 \omega{\omega'}^2}{(1-\omega^2 )^2} V_4 \varphi_0^2
\ea
and 
\be 
E=V_4 \tilde{U} + \frac{1}{4}\frac{Q^2}{V_4\varphi_0^2 }\, ,
\ee
where 
\be 
\label{tildeu}
\tilde{U} = U_5(\varphi_0^2 ) + \frac{P_5^2}{Q^2}\varphi_0^2.
\ee
Note that $P_5/Q = \omega' \omega /(1-\omega^2)$ so that
eq.(\ref{tildeu}) should be compared with
\be
\phi(x,y,t) = \varphi(x,y) \exp \left(
i \frac{P_5}{Q}(y+\frac{t}{\omega}) \right) .
\ee
Hence $\tilde{U}$ is the potential after applying the 
Scherk-Schwarz mechanism to the field $\phi$~\cite{ss}. One can now 
minimise the energy with respect to $V_4$ to find 
\be 
\label{energy}
E= Q\sqrt{\frac{\tilde{U}}{\varphi_0^2}}\, ,
\ee
where $\varphi_0 $ is the field value that minimises $E$.
Since the volume is proportional to $Q$, the point where $R$
becomes relatively small is determined by $Q$:
\be
Q \approx 4\pi R^4 \sqrt{\tilde{U}\varphi_0^2}.
\ee
Note that the potential $\hat{U}_{\omega\omega'}$ is 
the same for the large and small $R$ solutions (from now on we will drop the $\omega\omega'$ suffix),
and actually the energy is independent of $R$. 

We can see this in an interesting scaling limit of the thin wall 
approximation, given by 
\be
\omega' \ll 1 \qquad ; \qquad \omega^2 = 1 - {\cal O}(\omega')\, .
\ee
Without the intervention of $P_5$ small $\omega'$ would always imply small Q-balls,
but because of the simultaneous second limit we have 
\be 
Q \sim P_5 \sim V_4 \varphi_0^2 \, ,
\ee 
and at the same time $\hat{U}=U - {\cal O}(\omega')$, thereby maintaining the thin-wall requirement of  $\hat{U}(\varphi_0)\approx 0$.
 We conclude that in this limit the potential at $\varphi_0$  only needs to be shifted down by a parametrically small 
amount in order to develop a Q-ball solution, which nevertheless has large $Q$.
In more physical terms, the squared boost factor $1/(1-\omega^2) $ 
 is able to keep $Q$ and $P_5$ large even though $\omega'$ is small.
In this limit the energy is given by 
\be 
E= \frac{P_5^2}{Q^2} V \varphi_0^2+\frac{1}{4} \frac{Q^2}{V \varphi_0^2}+{\cal O}(\omega' )\, ,
\ee
which is minimised when 
\be 
V_{\rm min}=\frac{Q^2}{2P_5\varphi_0^2}+{\cal O}(\omega' ) \, .
\ee
We conclude that there is an energetically optimal volume for the Q-ball to occupy
given by the parameters on the right of this equation, but the Q-ball can achieve this minimal volume 
for \emph{any} radius of compact dimension $R$ because there is no surface tension term in the energy.
 Note that substituting back in we find that 
the energy of this configuration is $E=P_5+{\cal O}(\omega')$; i.e. the 4-dimensional 
rest-mass is made up almost entirely of $P_5$ in this limit, while the 5-dimensional rest-mass is negligible in this limit.

Returning to the generic case, we still need to show stability of the Q-ball with respect to decay 
into a collection of Kaluza-Klein modes. Decay is allowed into 
$Q$ Kaluza-Klein modes $i=1..Q$, with total momentum $\sum_i P_{5i} = 
P_5$, and total rest mass
\be  
E_{KK-modes}= \sum_i^Q \sqrt{\frac{1}{2}\mu^2 + P_{5i}^2},
\ee
where $\mu^2 = \partial^2 U/\partial \varphi^2 $.
A simple geometric argument shows that this expression is minimised 
when the $P_5$ momentum is equally distributed amongst the Kaluza-Klein 
modes, $P_{5i}=P_5/Q$. Stability therefore requires
\be 
E_{KK-modes} = Q \sqrt{\frac{1}{2}\mu^2 + \frac{P_5^2}{Q^2}} > 
E = Q\sqrt{\frac{ \tilde{U}}{\varphi_0^2}}
\ee
or 
\be 
\mu^2 > \frac{2 U}{\varphi_0^2} .
\ee
This is {\em precisely} the condition for a Q-ball to exist in the 
$4$ dimensional theory. Hence the Q-balls with additional integer global 
Q-charge can be simply understood as the 
Kaluza-Klein ladder of the lowest lying Q-ball\footnote{We should remark that our findings contradict those of ref.\cite{demir} which concluded 
that different stability conditions and types of Q-balls can result.
That analysis began with a decomposition of the 
action into Fourier modes, arriving at an infinite 
and intractable set of coupled differential equations 
for the Kaluza-Klein modes. However, the interactions among the 
different modes must be consistent with the fact that 
they come from higher dimensional interactions. Once this 
constraint is taken into account the stability condition must be 
as above. There is in effect one and only one kind of Q-ball however one chooses to  
squash it into extra dimensions, and at least in the large charge limit there are for example no special bounds on the mass per unit charge associated with 
the finite compactification radius.}. This can be trivially seen from Eqs.(\ref{tildeu},\ref{energy}) which gives
\be
E^2 (P_5)= E^2(0) + P_5^2 
\ee
so that in the thin wall limit the Kaluza-Klein momentum $P_5$ can be boosted away to leave the rest-mass of the Q-ball in a non-compact volume: the large Q-ball is blind to the compactness of the extra dimension.
The momentum $P_5$ is naturally decomposed into Kaluza-Klein modes as  
\be 
P_5=Q\left( p+\frac{n}{R}\right)
\ee
where $n$ is an integer parametrising the Kaluza-Klein tower, 
whilst the non-integer $p$ represents the 
Scherk-Schwarz phase with 
\be 
\phi(x,y+ 2\pi R) = e^{i p} \phi(x,y).
\ee
The interpretation of the phase $p$ is that it is the non-integer momentum per unit charge.

Similar solutions can be found for a global 
unitary symmetry as we shall later see for Q-balls on dielectric branes.
In the more general cases we have to replace the phase by a
time dependent unitary rotation, but the rest of the analysis 
will go through unchanged. 

\subsection{Generalization to $d$ compact dimensions} 

The treatment above can be straightforwardly extended to multi-dimensional compact flat spaces. Consider a toroidal compactification 
on an untilted torus with $d$ compact radii $R_a$ where $a=1\cdots d$. Then eq.(\ref{eww}) becomes 
\ba
\label{eww2}
\varepsilon _{\bomega,\omega'} &=& E + \bomega\cdot \left\{ 
\bPee - \int \dd ^3 x \dd^d y ~\Big{(} \partial_t \phi^* \nabla_d \phi +
\nabla_d \phi^* \partial_t \phi \Big{)}\right\} 
\nonumber \\
&& + \omega' \left(Q- \frac{1}{i} \int \dd^3  x \dd^d y ~
\phi^* \stackrel{\leftrightarrow}{\partial}_t\phi \right) \, ,
\ea
where  ${\boldsymbol \omega}=\left\{ \omega_a \right\} $ and ${ \boldsymbol P}$ are now $d$-vectors. 
The square in the kinetic terms is completed as 
\ba
\label{first}
\varepsilon _{\omega,\omega'}
&=& \int \dd ^3 x \dd^d y ~\bigg(
\left|\bomega \cdot \nabla_d \phi - \partial_t \phi + 
i \omega' \phi \right|^2
\nonumber \\
&& +(\delta_{ab}-\omega_a\omega_b )
\left( \partial_a \phi - \frac{i \omega_a \omega' }
{1-\bomega.\bomega}\phi \right) 
\left( \partial_b \phi^* + \frac{i \omega_b \omega' }
{1-\bomega.\bomega}\phi^* \right)
\nonumber\\
&&
+
\partial_i \phi^* \partial_i \phi +
\hat{U}(\phi\phi^*) \bigg)
+\bomega \cdot \bPee + \omega' Q\, ,
\ea
where 
\be 
\hat{U}(\phi\phi^*)
=U_{4+d}(\phi\phi^* ) - \frac{{\omega'}^2 }{(1-\bomega\cdot \bomega)}\phi\phi^*.
\ee
As long as  $\bomega \cdot \bomega < 1$ so that the ``dual metric'' is positive definite, the previous arguments 
go through unchanged, and the energy is minimised where 
\be 
\theta(y,t)= \frac{\omega'}{1-\bomega\cdot\bomega } (\bomega \cdot {\boldsymbol y} + t ) ,
\ee
and 
\be 
\varphi=\varphi ( \bomega\cdot \left( {\boldsymbol y}+\bomega t\right) ) .
\ee
Inserting these into the constraints with the large volume solution minimised at $\varphi_0$, explicitly gives 
\ba
Q &=& \frac{2\omega'}{1-\bomega \cdot \bomega} V_{3+d} \varphi_0^2 \nonumber \\
\bPee &=& \frac{2 \bomega{\omega'}^2}{(1-\bomega \cdot \bomega)^2 } V_{3+d} \varphi_0^2\, ,
\ea
with
\be 
E=V_{3+d} \tilde{U} + \frac{1}{4}\frac{Q^2}{V_{3+d}\varphi_0^2 }
\ee
where 
\be 
\label{tildeu2}
\tilde{U} = U(\varphi_0^2 ) + \frac{\bPee\cdot \bPee}{Q^2}\varphi_0^2.
\ee

\section{Small Q-balls in even smaller boxes}

\subsection{An exact solution}
In the previous section we saw that Q-balls in the  
large charge limit are energetically independent of 
the size of the compactification, and in 
the thin wall approximation it is only the 
total volume they occupy in the bulk that matters. 

In this section, we wish to get some idea of surface effects. We therefore turn to 
a logarithmic potential for which exact 
Q-ball solutions can be found in certain limits;
continuing with the definition $\phi=\varphi e^{i\theta}$, the particular $U(1)$ invariant 
potential of interest is
\be 
U_5 = \mu^2 \varphi^2 \left( 
1-\log \frac{\varphi^2}{\varphi_0^2 } \right) + {\cal{O}}(\varphi^n).
\ee
This potential is particularly interesting for studying surface effects 
because it admits exact Q-ball solutions whose `surfaces' constitute
the whole Q-ball, whatever the charge. It has found use in a limited number of 
related works in the past, most recently \cite{Copeland:2014qra}.

The last term in the potential, 
${\cal{O}}(\varphi^n)$ (where $n$ is some large power),
is added to lift the potential at large field 
values $\varphi \geqsim \sqrt{e} \varphi_0$
thereby ensuring 
that it satisfies the requirement that the origin be the 
global minimum. However, the modified potential for finding the 
Q-ball can be written, 
\be 
\hat{U}=   \mu^2 \varphi^2 \left( 
1-\log \frac{\varphi^2}{\varphi_0^{\prime 2} } 
\right) + {\cal{O}}(\varphi^n),
\ee
where 
\be 
\varphi_0'=e^{-\frac{\omega'^2}{\mu^2(1-\omega^2 )} } \varphi_0 .
\ee
Even for modest value of $\alpha $ the potential 
goes negative at field values that are exponentially smaller than the 
values at which  $\varphi^n$  dominates and
consequently, 
for the purposes of finding the Q-ball solution, the latter  is 
negligible.  

Neglecting this term allows one to solve the 
equations of motion in eq.(\ref{EoM}) by separation of variables. 
Of course the analysis regarding the phases of $\phi$ goes through as before, but 
now the solution for its modulus  $\varphi$ can be written, 
\be 
\varphi = \varphi_0'Y(y) \Pi_i X_i(x_i),
\ee
giving 
\be 
X_i = e^ {\frac{1}{2}(1-\mu^2 x_i^2)}
\ee
and 
\be 
\label{yeq}
 \frac{\ddot{Y}}{Y}= - \frac{\mu^2}{1-\omega^2 } \log Y ^2 .
\ee
The problem is reduced to the one dimensional 
task represented by this last equation. 
In the large radius limit it naturally just gives the expected Lorentz boosted version 
of the solutions in the $x_i$ directions,  
\be 
\label{isolated}
Y \rightarrow  \exp\left( {\frac{1}{2}-\frac{\mu^2}{2} \frac{1}{1-\omega^2} (y+\omega t)^2} \right)\, .
\ee
In more general cases it can easily be solved 
numerically imposing the boundary conditions of periodicity in 
$y\rightarrow y+2 \pi R$. Note that the typical width of the  Q-ball in
the $y$-direction, $\sqrt{1-\omega^2}/\mu$ has the expected Lorentz contraction. 

Some examples are shown in Figures 1a-e where the compactification 
radius is increased from $R\mu \approx \sqrt{(1-\omega^2)/2}$ to
$R\mu = 2 \sqrt{(1-\omega^2)/2}$. In figure 1a the 
 value of the radius corresponds to the `natural'
period; that is $Y(y)$ 
is oscillating close to $Y=1$ in the upturned potential $-\hat{U}$.
The oscillation period is monotonically increasing with amplitude so 
that (uniquely for this potential) there is a hard cut-off below which the solution is completely three dimensional:   {\em when $R\mu < \sqrt{(1-\omega^2)/2}$ there can be no solutions except the $O(3)$ symmetric 
trivial one}, $Y(y)= 1$. 

As the radius increases so does the amplitude of oscillation 
in order to maintain the correct periodicity.
Extending the oscillation period (\ie compactification radius) 
significantly, forces $Y$ to approach the origin of $\varphi $. 
In other words, the solution quickly collapses to the $O(4)$ symmetric one.

 At the radius $R\mu \geqsim 2 \sqrt{(1-\omega^2)/2}$, there are 
two available solutions. One is the isolated $O(4)$ 
symmetric configuration of figure 1e, and the other is the
doubled solution in figures 2a-b. The latter corresponds to 
$Y(y)$ oscillating twice in the period $2\pi R$. Further expansion 
of the radius causes the doubled solution  to  condense 
into two isolated Q-balls in the bulk. However, the doubled solution
of figure 2a is energetically unstable to decay into the 
single isolated Q-ball with the same charge and $P_5$. 
Similarly two completely isolated Q-balls of 
this type will coalesce into one large one\footnote{Note that these statements are all with the caveat that there are 
no relative phases between the Q-balls. In more general cases they can attract or repel 
and can exchange charge continuously \cite{Bowcock:2008dn}. One would expect this to be so with compact dimensions 
as well, although it would be of interest to extend the study of ref. \cite{Bowcock:2008dn} to this case.}.
Figures 2a,b show  two cases of interest,
the first being the solution with $Y\approx 1$ and the 
second the isolated solution in eq.(\ref{isolated}).

\begin{figure}[t]
\begin{pspicture}[](0,0)(10,5)
\epsfysize=2in
\epsffile{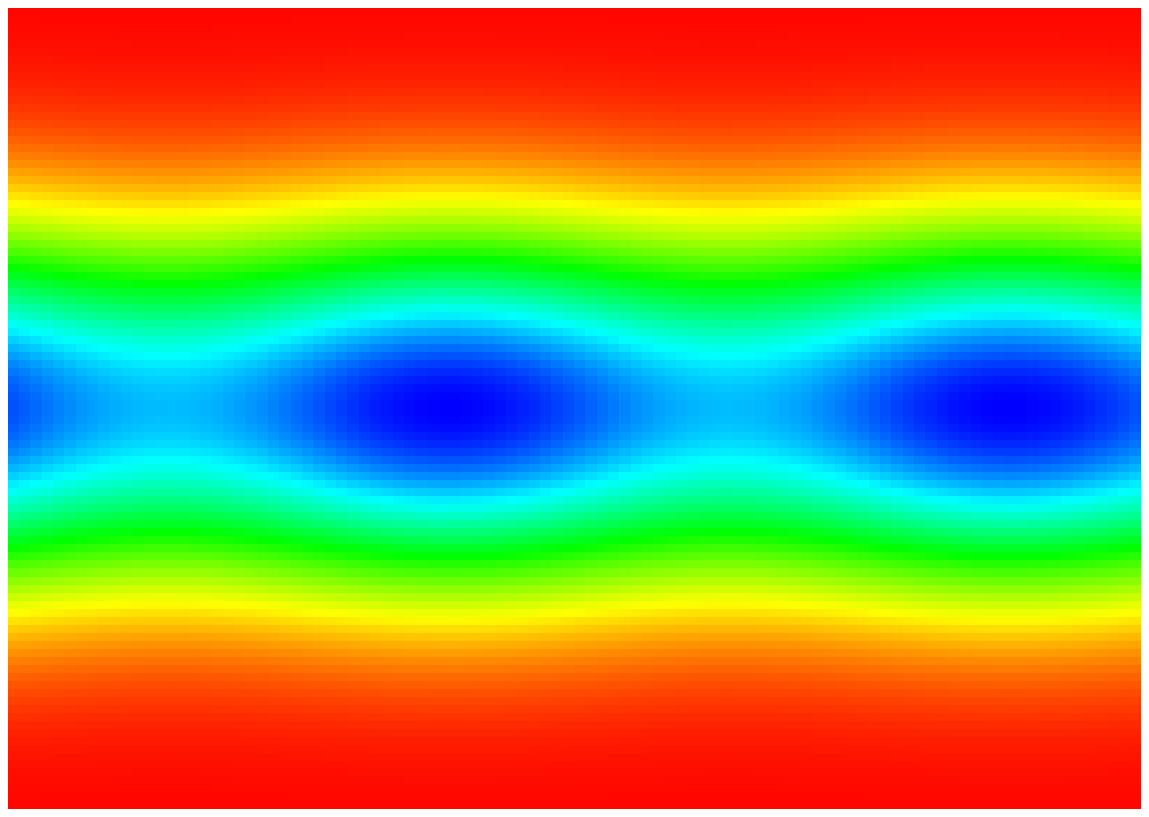}
\hspace{1.0cm}
\epsfysize=2in
\epsffile{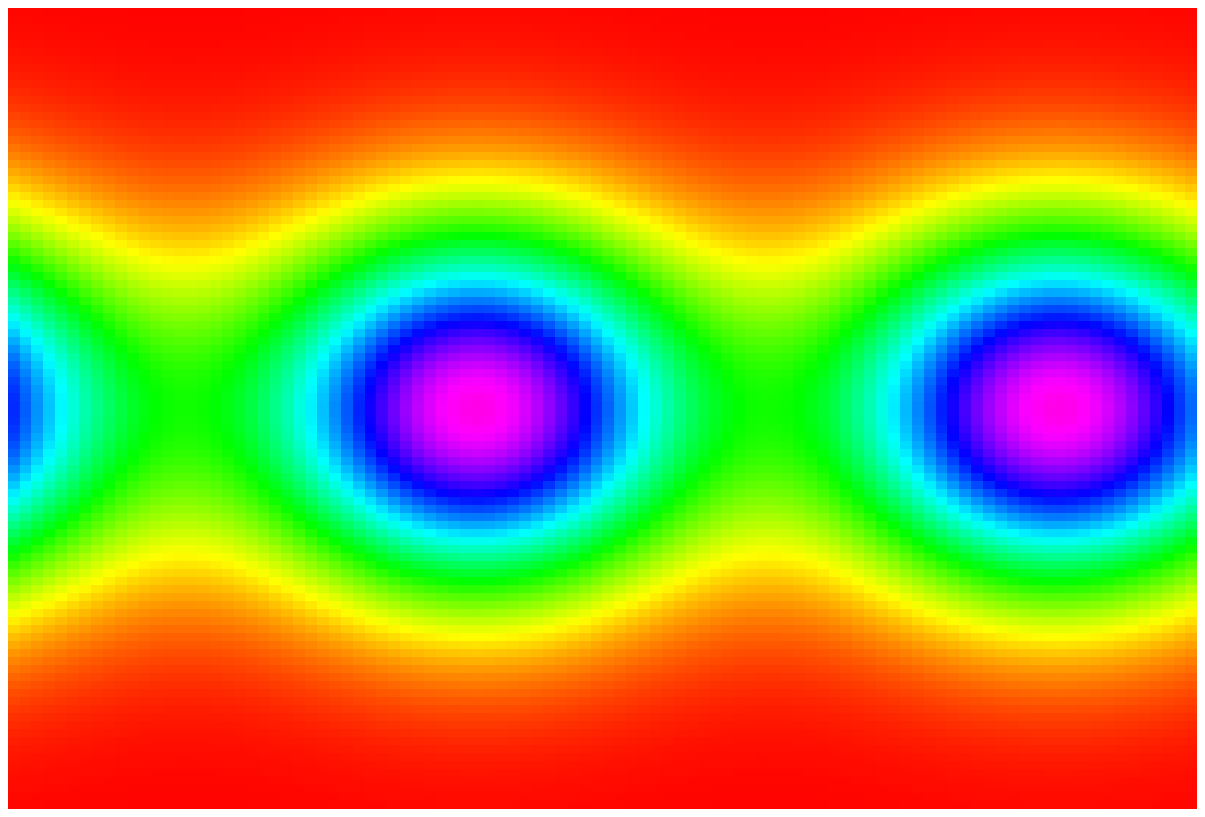}
\psline(0.00,0)(0.00,5.05)
\psline(-7.5,0)(-7.5,5.06)
\psline(-8.8,0)(-8.8,5.06)
\psline(-15.98,0)(-15.98,5.06)
\rput(-7.3,-0.4){b)}
\rput(-15.8,-0.4){a)}
\end{pspicture}
\vspace{0cm}
\caption{\it The second in a family of solutions first appearing at a radius that is twice 
the critical radius of the solutions in Figure 1. Increasing the 
radius of this solution results in the condensation of two 
isolated balls. Reducing it forces coalescence into one of the single Q-ball solutions 
in figure 1.} 
\end{figure}

We now present the charges, momentum and energy for these
different configurations.  To do so we define
\ba
\gamma &=& (e\sqrt{\pi})^3 \frac{\varphi^2_0 }{\mu^3 } \nn\\
     \xi &=& \frac{\omega'}{\mu\sqrt{1-\omega^2}}.
\ea
The general expressions are  (redefining $y+\omega t\rightarrow y$)  
\ba 
\label{51}
Q &=& 2\frac{\omega '}{1-\omega^2} \int \dd^3 x\dd y \varphi^2 \, 
\nonumber \\
&=& 2\frac{\omega '}{1-\omega^2} \gamma \int \dd y Y^2 \, , \nonumber \\
P_5 &=& \frac{\omega\omega'}{1-\omega^2} Q+ 2 \omega \int \dd^3x \dd y   (\partial_y \varphi)^2  \, 
\nonumber \\
   &=& \frac{\omega\omega'}{1-\omega^2} Q+ 2 \omega \frac{\mu^2}{1-\omega^2} \gamma \int \dd y Y^2  \log Y^2   \, ,
\ea
where in the last line we used 
eq.(\ref{yeq}) and integrated by parts for this example. 
This then gives for the squeezed and isolated limits respectively  \ba
Q_{sq} &=& \frac{2 R\mu \gamma  }{\sqrt{1-\omega^2}}  \xi e^{-\xi^2} \nn\\
Q_{is} &=& 2 e \sqrt{\pi} \gamma \xi e^{-\xi^2} \, .
\ea
This determines $\xi $ while $\omega $ can be determined by the equations for $P_5$; defining 
\be 
\omega = \frac{1}{\sqrt{1+\rho^2}} \, ,
\ee
the latter lead to
\ba 
\rho & = & \frac{Q\mu \xi }{P_5 } \nn\\
\rho & = & \frac{Q\mu \xi }{P_5 } (1+1/2\xi^2) \, ,
\ea 
and we can then parameterize the energy as  
\ba 
E_{sq} &=& Q_{sq} \mu \left( \frac{\alpha}{\mu}+\frac{1}{2}\frac{\mu}{\alpha} 
\right) \nn\\
E_{is} &=& Q_{is} \mu  \left( \frac{\alpha}{\mu}+\frac{1}{4}\frac{\mu}{\alpha} 
\right) \, ,
\ea
in the squeezed and isolated cases respectively, where as before $\alpha=\omega'/(1-\omega^2)$.

Notice that the minimum value for the energy 
of the isolated Q-ball, i.e. $Q\mu$, is less than that for the squeezed Q-ball $\sqrt{2}  Q\mu$, 
indicating that the effect of surface tension is for Q-balls energetically to favour large radius where they 
can assume a more symmetric configuration. 

To complete this discussion, we should remark that the Q-balls considered in this example are not unstable to decay into free 
states despite the fact that $E>  Q\mu$. This is because the parameter $\mu$ is not  the physical mass of any asymptotic quantum at the origin. (Formally the mass at the origin is logarithmically infinite so there are no asympotic states there at all.) 
Indeed consider the physical system in which Q-balls with such a potential could appear, namely the $F$ and $D$-flat directions 
corresponding to conserved $B-L$ current in supersymmetry, as was considered in for example ref.\cite{kusenko2}. The one-loop improved tree-level potential of this system would typically be of the form 
\be 
U = \mu^2 \varphi^2 \left( 
1-\log \frac{\varphi^2+m^2}{\varphi_0^2 } \right) + {\cal{O}}(\varphi^n),
\ee
where now $\varphi $ is the scalar denoting the VEV along the flat direction. The scale $\sqrt{\varphi^2+m^2}$ is the approximate renormalisation scale due to the field $\varphi$ giving a 
mass to for example squarks and sleptons along the flat direction, and the scale $m$ would therefore naturally 
correspond to the scale of supersymmetry breaking which would in  a typical supersymmetric phenomenology be of order $\mu$ itself. Provided 
$\varphi_0\gg m$ the Q-ball analysis goes through unchanged upto corrections of order ${\cal O}(m^2/\varphi_0^2)$, while the 
mass-squared of the asymptotic states at the origin, $\mu^2 \left( 1+\log \varphi_0^2/m^2 \right)$, 
is now regulated by the infra-red cut-off $m$, and is parametrically larger than $\mu^2$.

\subsection{The thick wall / small charge approximation}

We can also consider more general ``small'' Q-balls which would be more appropriate for the Q-balls on dielectric branes we discuss later. Following ref.\cite{kusenko} our task is to minimise
\ba
\varepsilon _{\omega,\omega'}
&=& \int \dd ^3 x \dd y ~ 
\bigg((\partial_y \varphi)^2 +
(\partial_i \varphi)^2 +
\hat{U}_{\omega\omega'}(\varphi^2)\bigg)
\ea
for fixed $\omega$ and $
\omega'$, 
where in the thick wall limit we keep only the first two terms in an expansion of the potential,
\be 
\hat{U}_{\omega\omega'}= \frac{\mu'^2}{2} \varphi^2 - A \varphi^3 + \ldots\, ,  \ee
  and the effective mass-squared  is 
\be 
\mu'^2 = \mu^2 - \frac{\omega'^2 }{1-\omega^2}\, .
\ee
Note that in a 5D theory, $\mu$ has mass-dimension 1, but $A$ has mass-dimension $1/2$.

The bounce action can be related by a simple rescaling to the bounce action for the rescaled  potential $V_\psi = \frac{1}{2}\psi^2 -\psi^3$ which can be computed numerically in certain cases \cite{kusenko}. The rescaling is of the form $\psi = \varphi A /\mu'^2 $ and $\hat{x}= \mu' x$, $\hat{y}=\mu' \sqrt{1-\omega^2} y $, so that the typical isolated solution would have $O(4)$ symmetry and width $\sim 1$ in the rescaled units, and again we infer squeezed solutions for $R \mu' \sqrt{1-\omega^2} < 1$. It is not possible to obtain the solution in full generality, however we can again restrict ourselves to either squeezed ($O(3)$ symmetric) or isolated ($O(4)$ symmetric) solutions as in ref.\cite{Linde:1981zj}. Considering the former for definiteness  gives $S_\psi = 4.85$ and an energy of 
\ba
\varepsilon _{\omega,\omega'} &=& 
S_\psi \frac{2\pi R \mu'^{3 }}{A^2} + \omega' Q + \omega P_5 \, .
\ea
Minimising in $\omega $ and $\omega'$ gives 
\ba
Q &=& \frac{ 3 S_\psi }{A^2} \frac{\omega'_{min}}{1-\omega_{min}^2} 2\pi R \mu_{min}' \nonumber \\ 
P_5 &=& \frac{\omega_{min} \,\omega'_{min}}{1-\omega_{min}^2}  Q\, ,
\ea
with of course the second relation following from eq.(\ref{51}). The energy can then be written 
\be 
E= Q\mu \left( \frac{\alpha}{\mu} +  \frac{1}{3}\frac{\mu}{\alpha} - \frac{1}{3} \frac{\omega'}{\mu}\right) \, .
\ee
The usual thick-wall solution of \cite{kusenko} has $w\equiv 0$ and hence $$E=Q\mu 
\left( \frac{2}{3}\frac{\omega'}{\mu} + \frac{1}{3}\frac{\mu}{\omega'} \right).$$ This gives $E> \frac{2\sqrt{2}}{3} Q\mu $ so the mass cannot be made {\em arbitrarily} small with respect to a collection of asymptotic quanta of the same charge. With non-zero $P_5$ a similar situation obtains, but with non-zero $\omega$ acting to increase the mass, such that 
\be 
E> \frac{2\sqrt{2+\omega^2}}{3} Q\mu . \label{eq0}
\ee 
We conclude that thick-wall Q-balls with $\omega >1/2$ are always unstable to decay.   

\section{Q-branes}
\subsection{Background: the dielectric brane potential}

We now turn to our particular application of the previous discussion, Q-balls as 
deformations of stacks of D$p$-branes. 

Let us briefly recap the Lagrangian for this system. As is well known, the massless modes of the   
open string form a supersymmetric $U(1)$ gauge theory with a vector 
$A_\mu, ~ \mu=0,1,..,p$, $9-p$ ``collective coordinate'' scalars $\Phi^i,~i=p+1,...,9$ and their 
fermionic partners. The dynamics of a single 
D$p$-brane is described by the DBI-action
\bea
S_{DBI}&=&-T_p\int d^{p+1}\sigma e^{-\phi_s}\sqrt{\det\Big{(}[G+
2\pi\alpha' B]_{\mu\nu}+2\pi\alpha'F_{\mu\nu}\Big{)}}\nonumber \\
&&\qquad\qquad\qquad\qquad \qquad\qquad+\mu_p\int \sum C^{(n)}\wedge e^{2\pi\alpha'(B+F)}\, ,
\label{DBI}
\eea
where $T_p, ~\mu_p$ are respectively the tension and the RR charge of the D$p$-brane,
$C^{(n)}$ is the $n+1$-form RR potential and $\phi_d$ is the string theory dilaton. We denote by $[..]$ the pull-back of spacetime tensors to the D$p$ worldvolume; for example
\begin{eqnarray}
[G]_{\mu\nu}=G_{\mu\nu}+4\pi \alpha' G_{i(\mu}\partial_{\nu)}\Phi^i+4\pi \alpha' G_{ij}\partial_\mu\Phi^i\partial_\nu\Phi^j. 
\end{eqnarray}
A collection of $N$ coincident
D$p$-branes supports a supersymmetric $U(N)$ gauge 
theory with  gauge field $A_\mu$, and scalars $\Phi^i$ in the 
adjoint of $U(N)$.  The latter act as the collective coordinates of the branes. The action which 
describes the dynamics  of such a collection of coincident D$p$-branes 
is not completely known. For example, 
replacing the abelian $U(1)$ in the action (\ref{DBI}) 
and taking the  symmetrized trace over the gauge group  as was 
suggested in \cite{Tseytlin} does not capture the full infrared dynamics
\cite{HT}, and in fact additional commutators of the field-strength are neeeded 
at sixth order \cite{Bain}.
Some progress can be made especially for the structure of the 
Chern-Simons term, the second term in (\ref{DBI}), in the non-abelian case
\cite{myers}. By using T-duality arguments, Myers showed that a D$p$-brane 
couples not only to the $p+1$-form RR potential but also to the RR potential with 
form degree higher than $p+1$ \cite{myers}. A collection of $N$ D0-branes for example in 
an electric  
RR four-form flux develops a dipole moment under the three-form potential.
 This is a ``dielectric'' property of the D$p$-branes   similar 
to the dielectric properties of neutral materials in  
electric fields. 
Indeed, in general, the Chern-Simons term for $N$ coincident D$p$-branes is 
modified to  \cite{myers}
\bea
S_{SC}=\mu_p\int \Tr\left(e^{i2\pi\alpha'i_\Phi i_\Phi}
\sum \Big{(}C^{(n)}\wedge e^{2\pi\alpha'(B+F)}\Big{)}\right)\, ,
\label{CS}
\eea
where $i_\Phi$ denotes the interior product by $\Phi^i$ if the latter is 
considered as a vector in the transverse space.  
The existence of these additional couplings in turn modifies the scalar potential 
of the world-volume theory. 
In the case of $N$ D$p$-branes,  for flat world-volume 
metric and vanishing RR and B-fields, the DBI-action, in lowest order in $\alpha'$,  turns out to be
\begin{eqnarray}
S_{DBI}=\int d^{p+1}\sigma \left[-4\pi^2\alpha'^2T_pe^{-\phi_s} \, \Tr\bigg(
D_\mu\Phi^i D^\mu \Phi^i+\frac{1}{4}F_{\mu\nu}F^{\mu\nu}\bigg)-V\right],
\end{eqnarray}
where the scalar potential is 
\bea
V=-T_p e^{-\phi_s}\pi^2\alpha'^2\Tr([\Phi^A,\Phi^B])^2\, .
\eea
Then by turning on an electric $p+3$-form potential $C_{01...pAB}$, 
an additional coupling of the D$p$-brane appears as can be seen from 
the Chern-Simons term (\ref{CS}),  so that 
the total potential turns out to be
\bea
\label{locpot}
 V=-T_pe^{-\phi_s}\pi^2\alpha'^2\Tr([\Phi^A,\Phi^B])^2-{i\over3} 
4\pi^2\alpha'^2\mu_p\Tr(\Phi^A\Phi^B\Phi^C)f_{ABC}\, , 
\eea  
where $f_{ABC}={1\over (p+1)!}\epsilon^{\mu_0...\mu_p}(\partial_A
C_{\mu_0...\mu_pBC}+cyclic).$

\subsection{Q-balls on dielectric branes}

Now let us consider simple Q-ball configurations  on such dielectric branes. 
For definiteness we will take $N$ coincident D3-branes; as the collective 
coordinate of the D3-branes plays the role of the (non-abelian) internal Q-charge, the
Q-lumps will describe the physical displacement of the D-branes within the compact dimensions, with the 
 D3-branes oriented so their internal Neumann dimensions fill space-time. 
The results extend trivially to other D$p$-branes. 

Generally speaking, non-abelian Q-ball solutions can be found in theories that have scalar fields $\phi_{ab}$  in a real $M\times M$ matrix representation of some non-abelian symmetry group \cite{SCA,AFK}. We should remark that the latter will turn out to be a subgroup of the $U(N)$ gauge symmetry described by the D$p$-branes so that the result will be gauged Q-balls rather than global. As discussed in \cite{Lee:1988ag} such objects are subject to a further constraint on their size coming from Coulomb repulsion of the charge which distributes itself over the surface of what is effectively a superconductor; however we will work in the small coupling limit in which this effect is negligible and the solutions are the same as the global ones. 

In the absence of gauge fields VEVs then, the 
action to lowest order after appropriate field redefinitions is 
\be 
S= \int \dd^{3+d} x \dd t \left(~\frac{1}{2}
(\partial \phi )^2- U(\phi)\right)\, ,
\ee
where $U(\phi)$ is the scalar potential, and traces over the $M\times M$ matrix indices are implied.
Generalising the results of the earlier sections, reparameterization invariance leads to the following 
conserved energies and momenta;
\ba
E &=& \int \dd ^{3+d} x ~\left(
\frac{1}{2}(\partial_t \phi )^2 +
\frac{1}{2}(\partial_i \phi )^2 + 
U(\phi )\right) \nonumber \\
P_i &=& \int \dd ^{3+d} x ~\partial_t \phi \partial_i \phi \, ,
\ea
where $i=1,\ldots 3+d$, again traces are inferred, and
we are assuming canonical kinetic terms. The conserved charges of the non-abelian symmetry are 
\be 
Q^k=\frac{1}{i} \int \dd^{3+d} x  ~ 
\partial_t \phi \left[ T^k , \phi\right] \, ,
\ee
where $T^k$ are the relevant generators.
To each charge we can associate a Lagrange multiplier, $\omega_k$, 
so that we must minimise the expression 
\be
\varepsilon_{\omega_k}  = E + \omega_k \left( 
Q^k - 
\frac{1}{i} \int \dd^{3+d} x  ~ 
\partial_t \phi \left[ T^k , \phi\right] \right)\, .
\ee
Completing the square and minimising as before 
we find 
\ba 
\phi & = & e^{i \omega_kT^k t } \varphi(x) \, e^{-i \omega_kT^k t } \nn\\
\varepsilon _{\omega_k}
&=& \int \dd ^{3+d} x  ~ \left(
\frac{1}{2}(\partial_i \varphi)^2 +
\hat{U}_{\omega_k}(\varphi) \right) 
+\omega_k Q^k ,
\ea
where 
\be 
\hat{U} = U + \frac{1}{2} Tr  \left( \omega_k 
\left[ T^k,\varphi \right] \right)^2 \, ,
\ee
and of course now $\varphi$ is also $M\times M$ matrix-valued.
Note that we could have found the same result by using the equations 
of motion, as we did for the $U(1)$ case discussed earlier. 

So far the discussion applies completely generally for non-abelian Q-balls. Our task now is to find a local minimum of the dielectric potential 
that preserves such a global non-abelian symmetry, and to determine $U(\phi)$ there. 
To do this let us turn on a background field,
\be 
 f_{ABC} =  f \varepsilon_{ABC}\, ,
\ee 
where we will take the $\Phi^A_{ab}$ to be  three  
$N\times N$ matrix-valued fields transforming under the $U(N)$, with $A=1..3$, $a,b = 1..N$.
As before, $A$ labels the three arbitrarily chosen extra dimensions in which we turn on the background field. 
As an ansatz, let the three fields $\Phi^A$ fall into an  irreducible 
$SU(2)$ multiplet as follows:
\begin{eqnarray}
\Phi^A(t,x)=\beta\, \hat{\phi}(t,x)\otimes \alpha^A\, ,
\end{eqnarray}
where the $\alpha^A $ form an $N/M\times N/M$ irreducible 
representation of $SU(2)$, $\hat{\phi}(t,x)$ is an arbitrary 
$M\times M$ 
real matrix and $\beta^{-1}=2\pi \alpha' T_p^{1/2}e^{-\phi_s/2}$ is a parameter that ensures canonical kinetic terms for $\hat \phi$. In particular we have that 
\begin{eqnarray}
[\alpha^A,\alpha^B]=2i\varepsilon_{ABC}\alpha ^C \, ,
\end{eqnarray}
and 
\begin{eqnarray}
{\rm Tr}(\alpha^A\alpha^B)=\frac{n}{3}(n^2-1)\delta^{AB}, ~~~n=N/M\, .
\end{eqnarray}
Inserting this ansatz into eq.(\ref{locpot}) and using the BPS condition for the tension and RR charge of the D$p$-branes, $T_p=\mu_p$, we find that ${V}$ becomes 
\begin{eqnarray}
V=8\pi^2 T_pe^{-\phi_s}\alpha'^2 \beta^4 n(n^2-1)\bigg(Tr\hat \phi ^4+\frac{\hat f}{3}Tr\hat \phi ^3\bigg)\, ,
\end{eqnarray}
where $\hat f= fe^{\phi_s}/\beta$. 
A local minimum exists at 
\begin{eqnarray}
\langle{\hat\phi}\rangle=\phi_0 I_{M}\, ,
\end{eqnarray}
where 
$\phi_0=-\hat f/4$ and $I_M$ is the $M\times M$ unit matrix.  This is the usual dielectric minimum, representing a configuration in which 
the $N$ D$3$-branes are bound to the surface of a D5-brane (forming a sphere in the non-space-time dimensions with radius $r_0=\frac{\pi}{2}\alpha' \hat f N$).

We may then define the D-brane displacements with respect to the shifted centres of mass of the blocks of $M$ D-branes as 
\begin{eqnarray}
\hat{\phi}(t,{x})=-\frac{\hat f}{4}I_{M}+\phi(t,{x})\, .
\end{eqnarray}
The $M\times M$ matrix-valued field $\phi(t,{x})$, corresponding to the displacement around the minimum, is precisely our 
desired non-abelian Q-ball field. Substituting into $V$ gives a potential for it of the form (ignoring a vacuum energy term)
\begin{eqnarray}
U(\phi)=\frac{1}{2}\mu^2{\rm Tr}\phi^2+\frac{g}{3!}{\rm Tr}\phi^3
+\frac{\lambda}{4!} {\rm Tr }\phi^4 \, ,\label{u1}
\end{eqnarray}
where
\begin{eqnarray}
\mu^2=\frac{\hat f^2\lambda}{96},~~~g=-\frac{\hat f\lambda}{6} ,\label{c1}
\end{eqnarray}
and 
\begin{eqnarray}
\lambda=\frac{12  n(n^2-1)e^{\phi_s}}{\pi^2T_p\alpha'^2}. \label{c2}
\end{eqnarray}
Note that $\lambda$ has engineering dimension $3-p$ as required for the action defined over the $p+1$ dimensional world volume. 
Clearly
eq.(\ref{u1}) is invariant under transformations
$\phi \rightarrow 
e^{i \Omega } \phi e^{-i \Omega }$, where $\Omega $ are elements 
of $GL(M,\mathbb{R})$, and we may therefore contemplate precisely the 
same Q-ball solutions as described above. 
Although the general case can be worked out, let us consider the  simplest case in which $\Omega $ are 
elements of $SO(3)$ (indeed minimal stable Q-balls are all unitarily equivalent to the $SO(3)$ Q-ball \cite{SCA}),
\be 
\phi(t,{x}) = e^{i \omega_k T_k t } \varphi({x})  
e^{-i \omega_k T_k t }\, ,
\ee
where $\varphi({x})$ is in the adjoint of $SO(3)$, and $T_k$ $(k=1,2,3)$ are $SO(3)$ generators in 
 the fundamental representation.
The potentials $U$ and $\hat{U}$ are shown in figure 3. This case was explicitly worked out in \cite{SCA,AFK}, with the result that the necessary and sufficient condition for the existence of Q-balls \cite{SCA} is 
\begin{equation}
1\leq \frac{g^2}{\mu^2\lambda}<9. \label{condition}
\end{equation}
The lower bound is the energetic condition for the existence of the Q-balls (ensuring  that 
the Q-ball will not decay into free mesons), whereas the upper bound is the condition 
that the cubic coupling is not very large so that $\phi=0$ is the global minimum. 
For the case at hand we have 
\begin{equation}
\frac{g^2}{\mu^2\lambda}=\frac{8}{3}\, ,
 \end{equation}
and thus eq.(\ref{condition}) is satisfied.  Therefore, dielectric branes support stable Q-balls in their world-volume. 

Given this, it is interesting to ask what their mass can be. Adopting the small Q-ball approximation, eq.(\ref{eq0}) gives $E > \frac{2\sqrt{2}}{3} Q\mu $ if we, for simplicity, take the lump to be non-relativistic in the compact dimensions, $\omega \ll 1$. In this limit our 5 dimensional system (which would correspond to dielectric D4 branes wrapped on a dimension of size $2\pi R$) would give Q-balls with a mass less than $Q \mu = {81}{\pi}S_\psi \frac{\omega' R}{\lambda}$, so the minimum Q-ball mass is proportional to the compactification radius measured in units of the Compton wavelength $1/\omega'$.  As we saw this number can in principle be less than unity. (Precisely how small it can be depends on the complicated dynamics of the Q-charge exchange which is beyond the scope of this paper to discuss, but would require further studies along the lines of \cite{Bowcock:2008dn,Copeland:2014qra}.) In addition $\lambda$ scales as $n^3$ and can therefore be large. We conclude that such fundamental Q-balls could be significantly  less massive than the fundamental scale. 
\begin{figure}[t]
\hspace{0in}
\epsfysize=12in
\epsffile{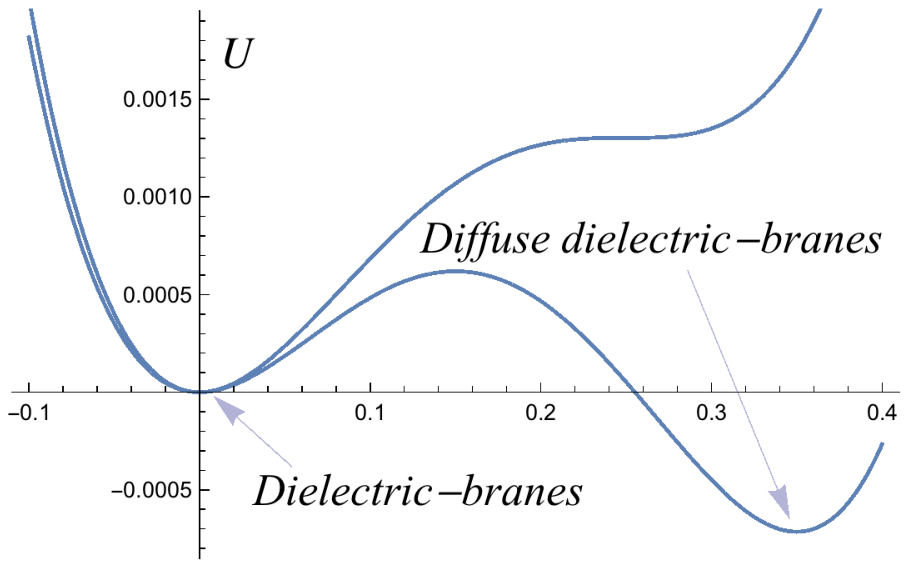}
\vspace{-22cm}
\caption{\it The original dielectric potential $U$ (upper), and the potential around its minimum $\hat{U}$ (lower) for 
non-abelian Q-ball deformations.} 
\end{figure}


\subsection*{Acknowledgements}
We would like to thank the Cern Theory division for hospitality and Paul Sutcliffe for useful discussions. This work was not supported by any military agency.


\begin{thebibliography}{99}

\bibitem{leepang}
For an early review see T.D.Lee and Y.Pang, ``Non-topological solitons'', \pr{221}{251}{1992}, and references therein.

\bibitem{coleman}
S.R.Coleman, ``Q-balls'', \npb{262}{263}{1985}

\bibitem{Lee:1988ag}
  K.~M.~Lee, J.~A.~Stein-Schabes, R.~Watkins and L.~M.~Widrow,
  ``Gauged Q-Balls,''
  Phys.\ Rev.\ D {\bf 39} (1989) 1665.

\bibitem{SCA}
  A.~M.~Safian, S.~R.~Coleman and M.~Axenides,
 ``Some Nonabelian Q Balls,''
  Nucl.\ Phys.\ B {\bf 297} (1988) 498.

\bibitem{kusenko}
  A.~Kusenko,
  ``Small Q balls,''
  Phys.\ Lett.\ B {\bf 404}, 285 (1997)
  [hep-th/9704073].

\bibitem{AFK}
  M.~Axenides, E.~Floratos and A.~Kehagias,
  ``NonAbelian Q balls in supersymmetric theories,''
  Phys.\ Lett.\ B {\bf 444} (1998) 190
  [hep-ph/9810230].

\bibitem{Arai:2008kv} 
  M.~Arai, C.~Montonen and S.~Sasaki,
  ``Vortices, Q-balls and Domain Walls on Dielectric M2-branes,''
  JHEP {\bf 0903}, 119 (2009)
  [arXiv:0812.4437 [hep-th]].
  
\bibitem{kusenko2}
For early references see
  A.~Kusenko,
  ``Phase transitions precipitated by solitosynthesis,''
  Phys.\ Lett.\ B {\bf 406}, 26 (1997)
  [hep-ph/9705361];
   A.~Kusenko and P.~J.~Steinhardt,
  ``Q ball candidates for selfinteracting dark matter,''
  Phys.\ Rev.\ Lett.\  {\bf 87}, 141301 (2001)
  [astro-ph/0106008];
  A.~Kusenko,
  ``Solitons in the supersymmetric extensions of the standard model,''
  Phys.\ Lett.\ B {\bf 405}, 108 (1997)
  [hep-ph/9704273];

\bibitem{star}
  B.~Hartmann and J.~Riedel,
  ``Supersymmetric Q-balls and boson stars in (d+1) dimensions,''
  Phys.\ Rev.\ D {\bf 87} (2013) 4,  044003
  [arXiv:1210.0096 [hep-th]].
\bibitem{star2}
  Y.~Brihaye, V.~Diemer and B.~Hartmann,
  ``Charged Q-balls and boson stars and dynamics of charged test particles,''
  Phys.\ Rev.\ D {\bf 89} (2014) 8,  084048
  [arXiv:1402.1055 [gr-qc]].

\bibitem{kusenko-dm}
A.~Kusenko and M.~E.~Shaposhnikov,
  ``Supersymmetric Q balls as dark matter,''
  Phys.\ Lett.\ B {\bf 418}, 46 (1998)
  [hep-ph/9709492];

\bibitem{Graham}
  P.~W.~Graham, S.~Rajendran and J.~Varela,
  ``Dark Matter Triggers of Supernovae,''
  arXiv:1505.04444 [hep-ph].

\bibitem{Kusenko:1997vp} 
  A.~Kusenko, V.~Kuzmin, M.~E.~Shaposhnikov and P.~G.~Tinyakov,
  ``Experimental signatures of supersymmetric dark matter Q balls,''
  Phys.\ Rev.\ Lett.\  {\bf 80}, 3185 (1998)
  [hep-ph/9712212].

\bibitem{Gia}
  G.~R.~Dvali, A.~Kusenko and M.~E.~Shaposhnikov,
  ``New physics in a nutshell, or Q ball as a power plant,''
  Phys.\ Lett.\ B {\bf 417} (1998) 99
  [hep-ph/9707423].

\bibitem{Yanagida}
  S.~Kasuya, M.~Kawasaki and T.~T.~Yanagida,
  ``IceCube potential for detecting Q-ball dark matter in gauge mediation,''
  PTEP {\bf 2015} (2015) 5,  053B02
  [arXiv:1502.00715 [hep-ph]].

\bibitem{japan2}
  J.~P.~Hong, M.~Kawasaki and M.~Yamada,
  ``Charged Q-balls in gauge mediated SUSY breaking models,''
  arXiv:1505.02594 [hep-ph].
\bibitem{arafuta}
  J.~Arafune, T.~Yoshida, S.~Nakamura and K.~Ogure,
  ``Experimental bounds on masses and fluxes of nontopological solitons,''
  Phys.\ Rev.\ D {\bf 62} (2000) 105013
  [hep-ph/0005103].

\bibitem{Takenaga}
  Y.~Takenaga {\it et al.}  [Super-Kamiokande Collaboration],
  ``Search for neutral Q-balls in super-Kamiokande II,''
  Phys.\ Lett.\ B {\bf 647} (2007) 18
  [hep-ex/0608057].

\bibitem{Ambrosio}
  M.~Ambrosio {\it et al.}  [MACRO Collaboration],
  ``Nuclearite search with the macro detector at Gran Sasso,''
  Eur.\ Phys.\ J.\ C {\bf 13} (2000) 453
  [hep-ex/9904031].

  \bibitem{myers}
R.~C.~Myers,
``Dielectric-branes,''
JHEP {\bf 9912}, 022 (1999), hep-th/9910053.

\bibitem{ss}
J. Scherk and J. H. Schwarz, ÒSpontaneous Breaking of Supersymmetry Through Dimensional Reduction,Ó Phys. Lett. 
{\bf B82} (1979)60.

\bibitem{demir} 
  D.~A.~Demir,
  ``Stable Q balls from extra dimensions,''
  Phys.\ Lett.\ B {\bf 495}, 357 (2000)
  [hep-ph/0006344].

\bibitem{Copeland:2014qra} 
  E.~J.~Copeland, P.~M.~Saffin and S.~Y.~Zhou,
  ``Charge-Swapping $Q$-balls,''
  Phys.\ Rev.\ Lett.\  {\bf 113}, no. 23, 231603 (2014)
  [arXiv:1409.3232 [hep-th]].

\bibitem{Bowcock:2008dn} 
 R.~Battye and P.~Sutcliffe,
  ``Q-ball dynamics,''
  Nucl.\ Phys.\ B {\bf 590} (2000) 329
  [hep-th/0003252];
  P.~Bowcock, D.~Foster and P.~Sutcliffe,
  ``Q-balls, Integrability and Duality,''
  J.\ Phys.\ A {\bf 42}, 085403 (2009)
  [arXiv:0809.3895 [hep-th]].
  
\bibitem{Linde:1981zj} 
  A.~D.~Linde,
  ``Decay of the False Vacuum at Finite Temperature,''
  Nucl.\ Phys.\ B {\bf 216}, 421 (1983)
  [Nucl.\ Phys.\ B {\bf 223}, 544 (1983)].
  
\bibitem{Tseytlin} A. Tseytlin, Nucl. Phys. {\bf B501} (1997) 41, 
hep-th/9701125; hep-th/9908105.

\bibitem{HT} A. Hashimoto and W. Taylor, Nucl. Phys. {\bf B503} 
(1997)193, hep-th/9703217.

\bibitem{Bain}   P.~Bain,
  ``On the nonAbelian Born-Infeld action,''
  hep-th/9909154.
  
\end{thebibliography}
\end{document}